\documentclass[12pt]{iopart}
\usepackage{graphicx}

\begin{document}

\title{Displacement deformed quantum fields}
\author{Peter Morgan}
\address{Physics Department, Yale University, CT 06520.}
\ead{peter.w.morgan@yale.edu}

\begin{abstract}
A displacement operator $\hat d_\zeta$ is introduced, verifying commutation relations
$[\hat d_\zeta, a_f^\dagger]=[\hat d_\zeta, a_f]=\zeta(f)\hat d_\zeta$ with field creation and
annihilation operators that verify $[a_f,a_g]=0$, $[a_f,a_g^\dagger]=(g,f)$, as usual.
$f$ and $g$ are test functions, $\zeta$ is a Poincar\'e invariant real-valued function on the
test function space, and $(g,f)$ is a Poincar\'e invariant Hermitian inner product.
The $\star$-algebra generated by all these operators, and a state defined on it, nontrivially extends
the $\star$-algebra of creation and annihilation operators and its Fock space representation.
If the usual requirement for linearity is weakened, as suggested in
\textsf{quant-ph/0512190}, we obtain a deformation of the free quantum field.
\end{abstract}

\maketitle

\newcommand\Half{{\frac{1}{2}}}
\newcommand\Intd{{\mathrm{d}}}
\newcommand\eqN{{\,\stackrel{\mathrm{N}}{=}\,}}
\newcommand\PP[1]{{(\hspace{-.27em}(#1)\hspace{-.27em})}}
\newcommand\PPs[1]{{(\hspace{-.4em}(#1)\hspace{-.4em})}}
\newcommand\RR {{\mathrm{I\hspace{-.1em}R}}}
\newcommand\CC{{{\rm C}\kern -0.45em 
          \vrule width 0.05em height 0.65em depth -0.03em
          \kern 0.45em}}
\newcommand\ZZ{{\mathrm{Z\hspace{-.4em}Z}}}
\newcommand\kT{{{\mathsf{k_B}} T}}
\newcommand\RE{{\mathrm{Re}}}
\newcommand\IM{{\mathrm{Im}}}

\newcommand\SF[2]{{\scriptstyle\frac{#1}{#2}}}
\newcommand\COMMENT[1]{{}}

\section{Introduction}
In an earlier paper, I introduced a weakening of the axioms of quantum field theory that allows a
nonlinear inner product structure \cite{MorganWLQF}. I refer to that paper for notation, motivation,
and an introduction to the approach that is further pursued here.
There, I mentioned that I had investigated deformations of the Heisenberg algebra of the Arik-Coons
type \cite{Quesne}, but had found no way to apply deformations of a comparable type to quantum fields.
Here, I briefly describe the failure, and move on to introduce a displacement operator
$\hat d_\zeta$, verifying $[\hat d_\zeta, a_f^\dagger]=[\hat d_\zeta, a_f]=\zeta(f)\hat d_\zeta$,
where $\zeta$ is an arbitrary real-valued scalar function on the test function space (taken to be
a Schwartz space \cite[\S II.1]{Haag}), which will allow us to construct an extension of
Fock space, generated by the action of displacement operators on a vacuum state as well as by the
action of creation operators $a_f^\dagger$.
Note that the ``displacement'' is not a space-time displacement, but will shortly be seen to ``displace''
creation and annihilation operators in the sense of adding a scalar.
What follows will show some of the uses to which such operators can be put.

A comparable (but Hermitian) number operator $\hat n_\zeta$ would verify the very different commutation
relation $[\hat n_\zeta, a_f^\dagger]=\zeta(f)a_f^\dagger$.
Number operators are important for a uniform presentation of algebras of the Arik-Coons type\cite{Quesne},
but we cannot in general construct an associative algebra if we use the operator $\hat n_\zeta$ to extend
the free quantum field algebra\,; it is straightforward to verify, for example, that for the undeformed
commutation relation $[a_f,a_g^\dagger]=(g,f)$, $\hat n_\zeta a_f a_g^\dagger$ becomes either
$(a_g^\dagger a_f+(g,f))(\hat n_\zeta-\zeta(f)+\zeta(g))$ or
$a_g^\dagger a_f(\hat n_\zeta-\zeta(f)+\zeta(g))+(g,f)\hat n_\zeta$, depending on the order in which
the commutation relations are applied, which is incompatible with associativity unless $\zeta$ is a
constant function on the test function space.
We will here take the constant function number operator to be relatively uninteresting, particularly
because we cannot generate an associative algebra using both a number operator $\hat n_1$ (with the
constant function $1$) and a displacement operator $\hat d_\zeta$\,; $\hat d_\zeta\hat n_1 a_f^\dagger$,
for example, becomes different values depending on the order in which commutation relations are applied.
Equally, every attempt I have made at deforming the commutation relations $[a_f,a_g^\dagger]=(g,f)$ and
$[a_f,a_g]=0$ using number operators or displacement operators have failed to be associative, with
$a_f(a_h a_g^\dagger)\not=a_h(a_f a_g^\dagger)$.

We will work with a $\star$-algebra $\mathcal{A}_1$ that is generated by creation and annihilation
operators that verify $[a_f,a_g^\dagger]=(g,f)$ and $[a_f,a_g]=0$, together with a single displacement
operator pair $\hat d_\zeta$ and $\hat d_\zeta^\dagger$.
We will take $\hat d_\zeta^\dagger$ to be equivalent to $\hat d_{-\zeta}$; $\hat d_\zeta^k$ to be
equivalent to $\hat d_{k\zeta}$; and $\hat d_{0\zeta}$ to be equivalent to $1$.
The commutation relations above and the state we will define in a moment are consistent with these
equivalences.
$\hat d_{0\zeta}$ is central in $\mathcal{A}_1$, for example.
In general, we will take $\hat d_{m\zeta}\hat d_{n\zeta}$ to be equivalent to $\hat d_{(m+n)\zeta}$.

$\mathcal{A}_1$ has the familiar subalgebra $\mathcal{A}_0$ that is generated by the creation and
annihilation operators alone.
A basis for $\mathcal{A}_1$ is
$a_{g_1}^\dagger a_{g_2}^\dagger...a_{g_m}^\dagger \hat d_{k\zeta} a_{f_1} a_{f_2}...a_{f_n}$, $k\in\ZZ$, for
some set of test functions $\{f_i\}$.
We construct a linear state $\varphi_0$ on this basis as
\begin{eqnarray}
  &&\varphi_0(1)=1,\\
  &&\varphi_0(a_{g_1}^\dagger a_{g_2}^\dagger...a_{g_m}^\dagger
                    \hat d_{k\zeta} a_{f_1} a_{f_2}...a_{f_n})=0
       \quad\mathrm{if}\ m>0\ \mathrm{or}\ n>0\ \mathrm{or}\ k\not=0.
\end{eqnarray}
If $k$ is always zero, this is exactly the vacuum state for the conventional free quantum field.
To establish that $\varphi_0$ is a state on $\mathcal{A}_1$, we have to show that
$\varphi_0(\hat A^\dagger \hat A)\ge 0$ for every element of the algebra.
A general element of the algebra can be written as
$\hat A=\sum_k\sum_r\lambda_{kr}\hat X^\dagger_{kr}\hat d_{k\zeta} \hat Y_{kr}$, where $\hat X_{kr}$
and $\hat Y_{kr}$ are products of annihilation operators, so that
\begin{eqnarray}
  \varphi_0(\hat A^\dagger \hat A)&=&\varphi_0(
      (\sum_j\sum_s\lambda_{js}^*\hat Y^\dagger_{js}\hat d_{-j\zeta} \hat X_{js})
      (\sum_k\sum_r\lambda_{kr}\hat X^\dagger_{kr}\hat d_{k\zeta} \hat Y_{kr}))\cr
  &=&\sum_k\varphi_0(
      (\sum_s\lambda_{js}^*\hat Y^\dagger_{js}{\hat X}'_{js})
      (\sum_r\lambda_{kr}\hat{X'}^\dagger_{kr}\hat Y_{kr}))\cr
  &=&\sum_k\varphi_0(\hat A_k^\dagger\hat A_k)\ge 0,
\end{eqnarray}
because only terms for which $j=k$ contribute, and
$\hat A_k=\sum_r\lambda_{kr}\hat{X'}^\dagger_{kr}\hat Y_{kr}$ is an operator in the free quantum field
algebra $\mathcal{A}_0$ for each $k$.
The critical observation is that ${\hat X}'_{kr}=\hat d_{-k\zeta}\hat X_{kr}\hat d_{k\zeta}$ is a sum of
products of annihilation operators only.

Given the state $\varphi_0$, we can use the GNS construction to construct a Hilbert space
$\mathcal{H}_0$ (see, for example, \cite[\S III.2]{Haag}), then we can use the $C^\star$-algebra of
bounded operators $\mathcal{B}(\mathcal{H}_0)$ that act on $\mathcal{H}_0$ as an algebra of observables,
but this or a similar construction is not strictly needed for \emph{Physics}.
From the point of view established in \cite{MorganWLQF}, we can be content to use a finite number
of creation operators and annihilation operators to generate a $\star$-algebra of operators.
This is not enough to support a continuous representation of the Poincar\'e group, but the
formalism is Poincar\'e invariant, adequate (if we take \emph{enough} generators) to construct
complex enough models to be as empirically adequate as a continuum limit, and is much simpler, more
constructive, and more appropriate for general use than Type $\mathrm{III}_1$ von Neumann algebras.
This paper broadly follows the general practice in physics of fairly freely employing unbounded
creation and annihilation operators.
Completion of a $\star$-algebra in a norm to give at least a Banach $\star$-algebra structure, which
would allow us to construct an action on the GNS Hilbert space directly, is a useful nicety for
mathematics, but it is not essential for constructing physical models.

For future reference, I list some of the simplest identities that are entailed by the
commutation relation of the displacement operator with the creation and annihilation operators
(using a Baker-Campbell-Hausdorff (BCH) formula for the exponentials):
\begin{eqnarray}
  &&[\hat d_\zeta^k,a_f^\dagger]=[\hat d_\zeta^k,a_f]=k\zeta(f)\hat d_\zeta^k,\\
 \label{DAnnihilation}
  &&\hat d_\zeta^k a_f^\dagger=(a_f^\dagger+k\zeta(f)) \hat d_\zeta^k,\qquad 
    \hat d_\zeta^k e^{i\lambda a_f^\dagger}=e^{i\lambda(a_f^\dagger+k\zeta(f))} \hat d_\zeta^k,\\
 \label{DCreation}
  &&\hat d_\zeta^k a_f=(a_f+k\zeta(f)) \hat d_\zeta^k,\qquad 
    \hat d_\zeta^k e^{i\lambda a_f}=e^{i\lambda(a_f+k\zeta(f))} \hat d_\zeta^k,\\
  &&e^{\alpha\hat d_\zeta-\alpha^*\hat d_\zeta^\dagger\,}a_f=
       \left[a_f+\zeta(f)(\alpha\hat d_\zeta+\alpha^*\hat d_\zeta^\dagger)\right]
            e^{\alpha\hat d_\zeta-\alpha^*\hat d_\zeta^\dagger}.
\end{eqnarray}
From these it should begin to be clear why I have called $\hat d_\zeta$ a ``displacement'' operator.
Equations (\ref{DAnnihilation}) and (\ref{DCreation}) make apparent the useful practical consequence
that it is sufficient to sum the powers of displacement operators in a term to be sure whether the term
contributes to $\varphi_0(\hat A)$ --- if the sum of powers is zero --- because displacement operators
are not modified if they are moved to left or right in the term.

We can introduce as many displacement operators as needed, all mutually commuting,
$[\hat d_{\zeta_1},\hat d_{\zeta_2}]=0$, without changing any essentials of the above, but probably
not as far as a continuum of such operators without significant extra care.
It is most straightforward to introduce linear dependency between products of the displacement
operators immediately, $\hat d_{\zeta_1}\hat d_{\zeta_2}=\hat d_{\zeta_1+\zeta_2}$, which is consistent
with the commutation relations, although we could also proceed by considering equivalence relations
later in the development.
The only other comment that seems necessary is that the action of the state $\varphi_0$ on a basis
constructed as above is zero unless there are no displacement operators present, so that
\begin{eqnarray}
  \varphi_0(1)=1,\quad\varphi_0(a_{g_1}^\dagger a_{g_2}^\dagger...a_{g_m}^\dagger
  \hat d_{\zeta_1}^{k_1}\hat d_{\zeta_2}^{k_2}...\hat d_{\zeta_l}^{k_l}
  a_{f_1} a_{f_2}...a_{f_n})&=&0,\cr
       &&\hspace{-10em}\quad\mathrm{if}\ m>0\ \mathrm{or}\ n>0\ \mathrm{or\ any}\ k_i\not=0.
\end{eqnarray}
$\hat d_{\zeta_1}^{k_1}\hat d_{\zeta_2}^{k_2}...\hat d_{\zeta_l}^{k_l}$ should be taken to be
equal to $\hat d_{k_1\zeta_1+k_2\zeta_2+\cdots+k_l\zeta_l}$.

The basic algebra is adequately defined above, the rest of this paper develops some of the
consequences for modelling correlations.
Three ways in which the displacement operators can be used are described below.
In particular, probability densities are calculated for various models, as far as possible.
All three ways can be combined freely with the two ways of constructing nonlinear quantum fields
that are described in \cite{MorganWLQF}, so the comment made there must be emphasized, that the
approach discussed here should at this point be considered essentially empirical, because there
is an embarrassing number of models.
The reason for pursuing this approach nonetheless --- from a high theoretical point of view the
lack of constraints on models might be seen as a serious failing --- is that it brings much better
mathematical control to discussions of renormalization, and might lead to new and hopefully useful
conceptualizations and phenomenological models of physical processes.
Even if the nonlinear quantum field theoretic models discussed here and in \cite{MorganWLQF} do
not turn out to be empirically useful, they nonetheless give an approach that can be compared
in detail with standard renormalization approaches, and an understanding of precisely why these
nonlinear models and others like them cannot be made to work should give some insight into both
approaches.

\section{Displaced vacuum states}
\label{VacuumDisplacement}
The way to use displacement operators that is discussed in this section in effect constructs
representations of the subalgebra $\mathcal{A}_0$, because the commutation relation
$[\hat\phi_f,\hat\phi_g]=(g,f)-(f,g)$ is unchanged.
However, we will be able to construct vacuum states in which the 1-measurement probability density in
the Poincar\'e invariant vacuum state can be any probability density in convolution with the
conventional Gaussian probability density, which seems useful regardless, particularly if used in
conjunction with the methods of \cite{MorganWLQF}.
The vacuum probability density may depend on any set of nonlinear Poincar\'e invariants of the test
function that describes a 1-measurement.

Let $\hat\phi_f=a_f+a_f^\dagger$ be the quantum field, for which the conventional vacuum state
generates a characteristic function $\chi_0(\lambda|f)$ of the 1-measurement probability density;
using a BCH formula, we obtain
\begin{eqnarray}
  \chi_0(\lambda|f)&=&\varphi_0(e^{i\lambda\hat\phi_f})=
    e^{-\Half\lambda^2(f,f)}\varphi_0(e^{i\lambda a_f^\dagger}e^{i\lambda a_f})\\
    &=&e^{-\Half\lambda^2(f,f)},
\end{eqnarray}
so that the probability density associated with single measurements in the vacuum state is
the Gaussian $\rho_0(x|f):=\exp{(-x^2/2(f,f))/\sqrt{2\pi(f,f)}}$.

Consider first the elementary alternative vacuum state,
$\varphi_d(\hat A)=\varphi_0(\hat d_\zeta \hat A \hat d_\zeta^\dagger)$.
For a vacuum state, $\zeta$ should be Poincar\'e invariant; this is a physical requirement on
vacuum states to which the mathematics here is largely indifferent.
Using this modified vacuum state, we can generate a characteristic function for single measurements,
\begin{eqnarray}
  \chi_d(\lambda|f)&=&\varphi_0(\hat d_\zeta e^{i\lambda\hat\phi_f} \hat d_\zeta^\dagger)=
    e^{-\Half\lambda^2(f,f)}\varphi_0(\hat d_\zeta e^{i\lambda a_f^\dagger}e^{i\lambda a_f} \hat d_\zeta^\dagger)\\
    &=&e^{-\Half\lambda^2(f,f)+2i\lambda\zeta(f)},
\end{eqnarray}
so that the probability density associated with single measurements in the modified vacuum state is
still Gaussian, but ``displaced'',
\begin{equation}
  \rho_d(x|f):=\frac{1}{\sqrt{2\pi(f,f)}}\exp{\left(-\frac{(x-2\zeta(f))^2}{2(f,f)}\right)}.
\end{equation}
As $\zeta(f)$ varies with some Poincar\'e invariant scale of $f$, the expected displacement of the Gaussian
varies accordingly.
$\zeta(f)$ might be large for ``small'' $f$, small at intermediate scale, and large again for ``large'' $f$;
any function of multiple Poincar\'e invariant scales of the test functions may be used.

Introducing a linear combination $\hat\Xi=\sum_k \xi_k\hat d_\zeta^k/\sqrt{N}$ of higher powers of
$\hat d_\zeta$, with normalization constant $N=\sum_k\left|\xi_k\right|^2$, we can construct another
modified vacuum state, $\varphi_c(\hat A)=\varphi_0(\hat\Xi \hat A \hat\Xi^\dagger)$, which generates
a characteristic function
\begin{eqnarray}
  \chi_c(\lambda|f)&=&\varphi_0(\hat\Xi e^{i\lambda\hat\phi_f} \Xi^\dagger)=
    e^{-\Half\lambda^2(f,f)}\varphi_0(\hat\Xi e^{i\lambda a_f^\dagger}e^{i\lambda a_f} \hat\Xi^\dagger)\\
    &=&\frac{1}{N}\sum_k \left|\xi_k\right|^2 e^{-\Half\lambda^2(f,f)+2ik\lambda\zeta(f)},
\end{eqnarray}
so that we obtain a probability density
\begin{equation}
  \rho_c(x|f)=\frac{1}{N}\sum_k \frac{\left|\xi_k\right|^2}{\sqrt{2\pi(f,f)}}
        \exp{\left(-\frac{(x-2k\zeta(f))^2}{2(f,f)}\right)}.
\end{equation}
If we are prepared to introduce a continuum of displacement operators, this probability density can
be any probability density in convolution with the conventional Gaussian probability density.
A finite number of displacement operators will generally be as empirically adequate as a continuum of
displacement operators.

Finally, we can explicitly generate the $n$-measurement probability density in the state
$\varphi_C(\hat A)=\varphi_0(\hat\Xi' \hat A \hat{\Xi'}^\dagger)$, where
$\hat\Xi'=\sum_m \xi'_m\hat d_{\zeta_m}/\sqrt{N'}$, with normalization constant
$N'=\sum_m\left|\xi'_m\right|^2$.
The characteristic function is
\begin{eqnarray}
  \chi_C(\lambda_1,\lambda_2,...,\lambda_n|f_1,f_2,...,f_n)&=&
   \varphi_0(\hat\Xi' e^{i\sum_j\lambda_j\hat\phi_{f_j}} {\Xi'}^\dagger)\\
    &=&\frac{1}{N'}\sum_m \left|\xi'_m\right|^2
          e^{-\Half\underline{\lambda}^T F\underline{\lambda}+2i\sum_j\lambda_j\zeta_m(f_j)},
\end{eqnarray}
where $F$ is the gram matrix $(f_i,f_j)$ and $\underline{\lambda}$ is a vector of the variables
$\lambda_i$.
$\chi_C(\lambda_1,\lambda_2,...,\lambda_n|f_1,f_2,...,f_n)$ generates the probability density
\begin{equation}
  \rho_C(x_1,x_2,...,x_n|f_1,f_2,...,f_n)=
        \frac{1}{N'}\sum_m \frac{\left|\xi'_m\right|^2}{\sqrt{2\pi \mathrm{det}(F)}}
        e^{-\Half\underline{x}(m)^T F^{-1}\underline{x}(m) },
\end{equation}
where the set of vectors $\underline{x}(m)$ is given by $x(m)_j=x_j-2\zeta_m(f_j)$.
With a suitable choice of $\zeta_m$ and $|\xi'_m|^2$, we can make the probability density vary with
multiple Poincar\'e invariant scales of the individual measurements.
Note, however, that in the approach of this paper only the gram matrix $F$ describes the relationships
\emph{between} the measurements described by the test functions $f_i$, and all such relationships are
pairwise.

\section{Displacements of the field observable-I}
\label{FieldDeformationI}
This and the following section introduce deformations of the field instead of deformations of
the ground state.
As above, the quantum field discussed in this section still satisfies the commutation relation
$[\hat\phi_f,\hat\phi_g]=(g,f)-(f,g)$, so the states we can construct again effectively
generate many representations of the free field algebra of observables (the \emph{next} section
modifies the commutation relations satisfied by the observable field).
If we think of ourselves as constructing empirically effective models for physical situations,
it is worth considering different models for the different intuitions they present, while
of course also presenting, as clearly as possible, isomorphisms between models, or -- less
restrictively -- empirical equivalences between models.

The simplest deformation discussed in this section is
\begin{equation}
  \hat\phi_f=i(a_f-a_f^\dagger)+\alpha(f)\hat d_\zeta+\alpha^*(f)\hat d_\zeta^\dagger,
\end{equation}
This deformed field satisfies microcausality because $\hat d_\zeta$ commutes with
$i(a_f-a_f^\dagger)$\footnote{Another possibility,
$\hat\phi'_f=a_f+a_f^\dagger+\zeta(f)(\alpha\hat d_\zeta+\alpha^*\hat d_\zeta^\dagger)$, also
satisfies microcausality, but is almost trivially seen to be unitarily equivalent to $a_f+a_f^\dagger$,
\begin{equation}
  e^{\Half(\alpha\hat d_\zeta-\alpha^*\hat d_\zeta^\dagger)}(a_f+a_f^\dagger)
      e^{-\Half(\alpha\hat d_\zeta-\alpha^*\hat d_\zeta^\dagger)}=
      a_f+a_f^\dagger+\zeta(f)(\alpha\hat d_\zeta+\alpha^*\hat d_\zeta^\dagger).
\end{equation}
This establishes a close enough relationship to the previous section that a longer presentation
of this case will not be given here.}.
Note that in this section and in the next we take $a_f+a_f^\dagger$ not to be an observable of the
theory, because $[(a_f+a_f^\dagger),i(a_g-a_g^\dagger)]\not=0$ when $f$ and $g$ have space-like
separated supports.

\begin{figure}[htb]
\centerline{\includegraphics[width=20em,height=20em]{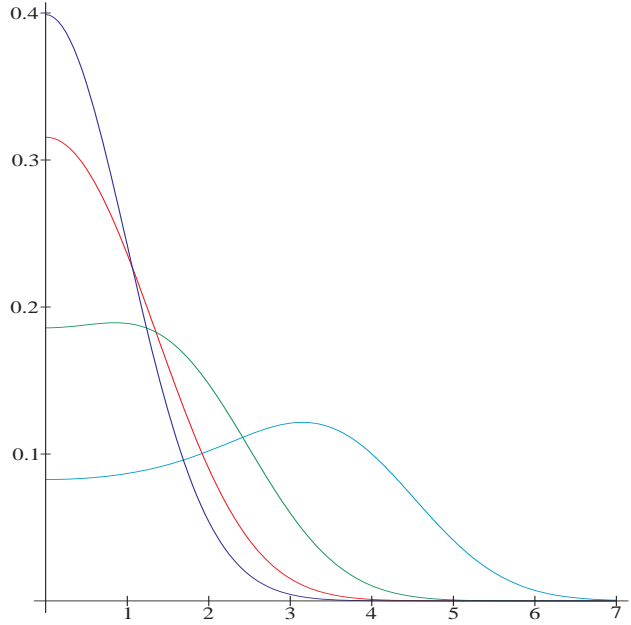}}
\caption{The probability densities that result from the deformation\\
  $\hat\phi_f=i(a_f-a_f^\dagger)+\alpha(f)\hat d_\zeta+\alpha^*(f)\hat d_\zeta^\dagger$, with
  $(f,f)=1$ and $|\alpha(f)|=0$ (blue,\\ highest function at zero),
                $\frac{1}{3}$ (red, second highest),
                $1$ (green, third\\ highest),
                $3$ (cyan, lowest function at zero) [colour on the web].}
\label{PhiDeformationOne}
\end{figure}

We can straightforwardly calculate the vacuum state 1-measurement characteristic function
for $\hat\phi_f$,
\begin{eqnarray}
  \chi_J(\lambda|f)&=&\varphi_0(e^{i\lambda\hat\phi_f})=
    e^{-\Half\lambda^2(f,f)}\varphi_0(e^{\lambda a_f^\dagger}e^{-\lambda a_f}
           e^{i\lambda(\alpha(f)\hat d_\zeta+\alpha^*(f)\hat d_\zeta^\dagger)})\cr
    \rule[-3ex]{0pt}{7ex}
    &=&e^{-\Half\lambda^2(f,f)}\sum_{j=0}^\infty \frac{(i\lambda|\alpha(f)|)^{2j}}{(2j)!}\frac{(2j)!}{j!^2}
                                          \varphi_0(e^{\lambda a_f^\dagger}e^{-\lambda a_f})\cr
    &=&e^{-\Half\lambda^2(f,f)}J_0(2\lambda|\alpha(f)|),
\end{eqnarray}
where the Bessel function emerges because the only contributions to the result are those for which
$\hat d_\zeta$ and $\hat d_\zeta^\dagger$ cancel, which gives the contribution $\frac{(2j)!}{j!^2}$.
This results in a probability density that is the convolution of the conventional Gaussian and
the probability density $\frac{1}{\sqrt{|2\alpha(f)|^2-x^2}}$ (when $|x|<|2\alpha(f)|$, otherwise $0$).
The probability density we have just calculated is independent of $\zeta$, because
$\hat d_\zeta$ commutes with $i(a_f-a_f^\dagger)$, but $\zeta$ will turn up in expressions for
non-vacuum state probability densities.
The scales of $(f,f)$ and $|\alpha(f)|$ determine the ``shape'' of the convolution.
The convolution is displayed in figure \ref{PhiDeformationOne} for $(f,f)=1$ and $|\alpha(f)|=0$,
$\frac{1}{3}$, $1$, and $3$.

We can also compute characteristic functions for higher powers such as
$\hat \phi_f=i(a_f-a_f^\dagger)+\alpha(f)(\hat d_\zeta+\hat d_\zeta^\dagger)^k$,
\begin{eqnarray*}
  k=1&\longrightarrow&{}_0F_1(;1;-(\lambda\alpha(f))^2)e^{-\Half\lambda^2(f,f)}
                      =J_0(2\lambda\alpha(f))e^{-\Half\lambda^2(f,f)},\\
  k=3&\longrightarrow&{}_2F_3(\SF{1}{6}, \SF{5}{6};
                              \SF{1}{3}, \SF{2}{3}, 1;-16(\lambda\alpha(f))^2)
                                e^{-\Half\lambda^2(f,f)},\\
  k=5&\longrightarrow&{}_4F_5(\SF{1}{10}, \SF{3}{10}, \SF{7}{10}, \SF{9}{10};
                              \SF{1}{5}, \SF{2}{5}, \SF{3}{5}, \SF{4}{5}, 1;-256(\lambda\alpha(f))^2)
                                e^{-\Half\lambda^2(f,f)},\\
  &\mathit{etc.},&\cr
  k=0&\longrightarrow&{}_0F_0(;;2i\lambda\alpha(f))e^{-\Half\lambda^2(f,f)}
                      =e^{2i\lambda\alpha(f)}e^{-\Half\lambda^2(f,f)},\\
  k=2&\longrightarrow&{}_1F_1(\SF{1}{2};1;4i\lambda\alpha(f))e^{-\Half\lambda^2(f,f)}
                      =J_0(2\lambda|\alpha(f)|)e^{2i\lambda\alpha(f)}e^{-\Half\lambda^2(f,f)},\\
  k=4&\longrightarrow&{}_2F_2(\SF{1}{4},\SF{3}{4};\SF{1}{2},1;16i\lambda\alpha(f))e^{-\Half\lambda^2(f,f)},\\
  k=6&\longrightarrow&{}_3F_3(\SF{1}{6},\SF{3}{6},\SF{5}{6};
                              \SF{1}{3},\SF{2}{3},1;64i\lambda\alpha(f))e^{-\Half\lambda^2(f,f)},\\
  &\mathit{etc.}& 
\end{eqnarray*}
The $k=0$ entry is trivially tractable, indeed trivial; otherwise only the $k=2$ entry is immediately
tractable, being just a trivially displaced version of the $k=1$ entry we have just discussed, because
$(d_\zeta+d_\zeta^\dagger)^2=(d_{2\zeta}+d_{2\zeta}^\dagger)+2$.
The combinatorics for arbitrary Hermitian functions of $\hat d_\zeta$ and $\hat d_\zeta^\dagger$ added
to $i(a_f-a_f^\dagger)$, potentially using multiple Poincar\'e invariant displacement functions $\zeta_i$,
can be as complicated as we care to consider.

Further possibilities that must be considered, because $\hat d_\zeta$ cannot generally be taken to be
linear in $\zeta$, are fields such as
$i(a_f-a_f^\dagger)+\alpha(f)(\hat d_{\beta(f)\zeta}+\hat d_{\beta(f)\zeta}^\dagger)$, which are distinct
from the other fields considered in this section even though the vacuum state 1-measurement probability
densities are independent of $\beta(f)\zeta$.
If we add two displacement function components, as in
$i(a_f-a_f^\dagger)+\alpha_1(f)(\hat d_{\beta_1(f)\zeta}+\hat d_{\beta_1(f)\zeta}^\dagger)
                   +\alpha_2(f)(\hat d_{\beta_2(f)\zeta}+\hat d_{\beta_2(f)\zeta}^\dagger)$
there is a complex modulation of the vacuum state 1-measurement probability density as the proportion
of $\beta_1(f)$ to $\beta_2(f)$ changes.

\section{Displacements of the field observable-II}
\label{FieldDeformationII}
The first deformation of $\hat\phi_f$ that we will discuss in this section is
\begin{equation}
  \hat\phi_f=i(a_f-a_f^\dagger)(\hat d_\zeta+\hat d_\zeta^\dagger).
\end{equation}
As in the previous section, this is Hermitian and satisfies microcausality, but the
algebra of observables generated by the observable field is finally different,
\begin{equation}
  [\hat\phi_f,\hat\phi_g]=[(g,f)-(f,g)](\hat d_\zeta+\hat d_\zeta^\dagger)^2,
\end{equation}
even though the algebra satisfied by the creation and annihilation operators is unchanged.
The change in the algebra of observables gives some cause to think that physics associated
with this type of construction may be significantly different.
$(\hat d_\zeta+\hat d_\zeta^\dagger)^2$ is a central element in the algebra
generated by $\hat\phi_f$.

\begin{figure}[htb]
\centerline{\includegraphics[width=20em,height=20em]{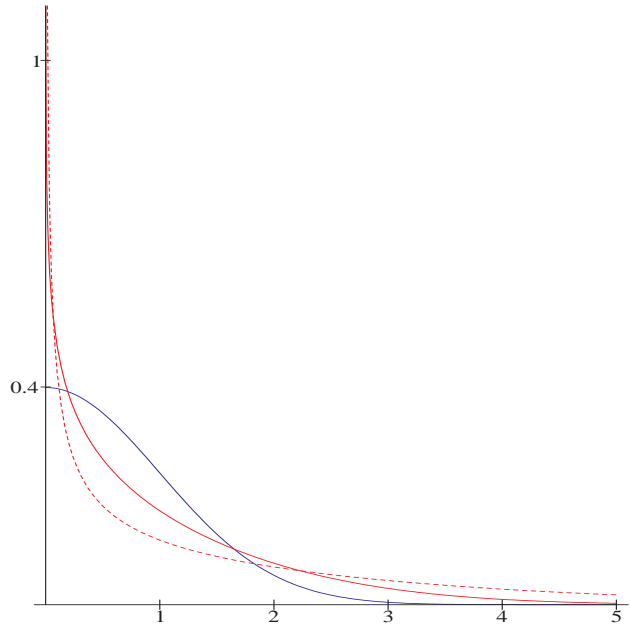}}
\caption{The probability density that results from the deformation\\
  $\hat\phi_f=i(a_f-a_f^\dagger)(\hat d_\zeta+\hat d_\zeta^\dagger)$, with
  $(f,f)=1$, variance 2 (in red), compared with\\ the conventional Gaussian,
  with $(f,f)$, variance 1 (in blue), and the probability\\ density
  that results from the deformation $\hat\phi_f=i(a_f-a_f^\dagger)(\hat d_\zeta+\hat d_\zeta^\dagger)^2$,
  with\\ $(f,f)=1$, variance 6 (dashed, in red)[colour on the web].}
\label{PhiDeformationTwo}
\end{figure}

The characteristic function of the vacuum state 1-measurement probability density is
\begin{eqnarray}
  \chi_P(\lambda|f)&=&\varphi_0(e^{i\lambda\hat\phi_f})\cr
    &=&\varphi_0\left(\sum_{j=0}^\infty \frac{(i\lambda)^j i^j (a_f-a_f^\dagger)^j
          (\hat d_\zeta+\hat d_\zeta^\dagger)^j}{j!}\right)\cr
    \rule[-4ex]{0pt}{9ex}
    &=&\varphi_0\left(\sum_{j=0}^\infty \frac{\lambda^{2j}(a_f-a_f^\dagger)^{2j}}{(2j)!}
                 \frac{(2j)!}{j!^2}\right)\cr
    &=&\sum_{j=0}^\infty \frac{(-\lambda^2(f,f))^j}{(2j)!}\frac{(2j)!}{2^j j!}\frac{(2j)!}{j!^2}\cr
    &=&{}_1F_1(\SF{1}{2};1;-2\lambda^2(f,f))=I_0(\lambda^2(f,f))e^{-\lambda^2(f,f)},
\end{eqnarray}
where $\varphi_0((a_f-a_f^\dagger)^{2j})=(-(f,f))^j\frac{(2j)!}{2^j j!}$ is a useful identity for the
conventional vacuum state.
$\chi_P(\lambda|f)$ can be inverse Fourier transformed, using \cite[\textbf{7.663}.2 or \textbf{7.663}.6]{GR},
to obtain
\begin{equation}\label{Keqn}
  \rho_P(x|f)=\frac{1}{\sqrt{8\pi^3(f,f)}}\exp{\left(-\frac{x^2}{16(f,f)}\right)}K_0\left(\frac{x^2}{16(f,f)}\right).
\end{equation}
This has variance $2(f,f)$, in contrast to the variance $(f,f)$ for the quantum field $i(a_f-a_f^\dagger)$.
$\rho_P(x|f)$ is displayed with variance $2(f,f)=2$ together with the Gaussian for $(f,f)=1$ in figure
\ref{PhiDeformationTwo}.
The vacuum state probability density $\rho_P(x|f)$ is again independent of $\zeta$; it is infinite at
zero, but it is also integrable enough over the real line for all finite moments to exist, which of
course we computed explicitly in order to compute $\chi_P(\lambda|f)$.

The probability density $\rho_P(x|f)$ is significantly concentrated both near zero and near
$\pm\infty$, relative to the conventional Gaussian probability density.
If we compare with a Gaussian that has the same variance, there is a 10 times greater probability
of observing a value beyond about 3.66 standard deviations, a 100 times greater probability of observing
a value beyond about 4.84 standard deviations, and a 1000 times greater probability of observing a value
beyond about 5.76 standard deviations.
I suppose $\rho_P(x|f)$ will give a fairly distinctive signature in physics, which future papers will
hopefully be able to make evident, and it should be clear fairly quickly whether it can be used to model
events in nature.

The characteristic function of the vacuum state $n$-measurement probability density is
\begin{equation}
  \chi_P(\lambda_1,\lambda_2,...,\lambda_n|f_1,f_2,...,f_n)=
         \varphi_0(e^{i\sum_j\lambda_j\hat\phi_{f_j}})=
         {}_1F_1(\SF{1}{2};1;-2\underline{\lambda}^T F\underline{\lambda}),
\end{equation}
where, as in section \ref{VacuumDisplacement}, $F$ is the gram matrix $(f_i,f_j)$ and
$\underline{\lambda}$ is a vector of the variables $\lambda_i$.
For $n=2$, we can inverse Fourier transform this radially symmetric function\footnote{Recall that
the $n$-dimensional inverse Fourier transform of a radially symmetric function $\tilde f(\rho)$ is given by
\begin{equation}
  \frac{1}{(2\pi)^{\frac{n}{2}}r^{\frac{n}{2}-1}}\int_0^\infty \tilde f(\rho)
                               \rho^{\frac{n}{2}}J_{\frac{n}{2}-1}(r\rho)\Intd\rho.
\end{equation}} using
\cite[\textbf{7.663}.5]{GR}, to obtain
\def\XX{{\underline{x}^T F^{-1}\underline{x}}}
\begin{equation}\label{Weqn}
  \rho_P(x_1,x_2|f_1,f_2)=
             \frac{\exp{\left(-\frac{\XX}{8}\right)}}
                  {\sqrt{8\pi^3(\XX)\mathrm{det}(F)}},
\end{equation}
For all $n$, we can confirm, using \cite[\textbf{7.672}.2]{GR} that the Fourier transform of
\begin{equation}
  \rho_P(x_1,x_2,...,x_n|f_1,f_2,...,f_n)=
    \frac{\exp{\left(-\frac{\XX}{16}\right)}
          W_{\frac{n}{4}-\frac{1}{4},\frac{n}{4}-\frac{1}{4}}\left(\frac{\XX}{8}\right)}
         {2^{\frac{3n}{4}-\frac{3}{4}}(\XX)^{\frac{n}{4}+\frac{1}{4}}\sqrt{\pi^{n+1}\mathrm{det}(F)}}
\end{equation}
is ${}_1F_1(\SF{1}{2};1;-2\underline{\lambda}^T F\underline{\lambda})$, where $W_{a,b}(z)$ is
Whittaker's confluent hypergeometric function.
Although these mathematical derivations of probability densities can be derived, and give a distinct insight,
the moments, which are essentially what are physically measurable, can be determined more easily from the
characteristic functions, or directly from the action of a state on an observable.

We can also compute characteristic functions for higher powers of displacement operators,
$\hat\phi_f=i(a_f-a_f^\dagger)(\hat d_\zeta+\hat d_\zeta^\dagger)^k$,
\begin{eqnarray*}
  k=1&\longrightarrow&{}_1F_1(\SF{1}{2};1;-2\lambda^2(f,f))
                      =I_0(\lambda^2(f,f))e^{-\lambda^2(f,f)},\\
  k=2&\longrightarrow&{}_2F_2(\SF{1}{4}, \SF{3}{4};
                              \SF{1}{2}, 1;-8\lambda^2(f,f)),\\
  k=3&\longrightarrow&{}_3F_3(\SF{1}{6}, \SF{3}{6}, \SF{5}{6};
                              \SF{1}{3}, \SF{2}{3}, 1;-32\lambda^2(f,f)),\\
  k=4&\longrightarrow&{}_4F_4(\SF{1}{8}, \SF{3}{8}, \SF{5}{8}, \SF{7}{8};
                              \SF{1}{4}, \SF{2}{4}, \SF{3}{4}, 1;-128\lambda^2(f,f)),\\
  &\mathit{etc.,}& 
\end{eqnarray*}
which in general have Meijer's $G$-functions as inverse Fourier transforms \cite[\textbf{7.542}.5]{GR}.
For $k=2$, again using \cite[\textbf{7.672}.2]{GR}, with different substitutions, we can derive the
probability density
\begin{equation}
  \rho_{P2}(x|f)=\frac{1}{\sqrt{64\pi^3(f,f)}}\exp{\left(-\frac{x^2}{64(f,f)}\right)}
                       K_{\frac{1}{4}}\left(\frac{x^2}{64(f,f)}\right),
\end{equation}
This has variance $6(f,f)$; it is plotted for $(f,f)=1$ in Figure \ref{PhiDeformationTwo}. 
In general we can multiply $i(a_f-a_f^\dagger)$ by any self-adjoint polynomial in $\hat d_{\beta(f)\zeta}$ and
$\hat d_{\beta(f)\zeta}^\dagger$.
It will be interesting to discover what range of probability densities this will allow us to construct.

\section{Discussion}
This mathematics is essentially quite clear and simple, but it is also rather rich
and nontrivial, and there are lots of concrete models.
It will be apparent that I do not have proper control of the full range of possibilities.
From philosophical points of view that seek a uniquely preferred model and that find the tight
constraints of renormalization on acceptable physical models congenial, it will be seen as problematic
that there is a plethora of models, but a loosening of constraints accords well with our experience
of wide diversity in the natural world, and is no more than a return to the almost unconstrained
diversity of classical particle and field models.

It is so far rather unclear how to understand the mathematics as physics, but any interpretation
will follow a common (but not universal) quantum field theoretical assumption that we measure
probabilities and correlation functions of scalar observables that are indexed by test functions.
There are existing ways of discussing condensed matter physics that are fairly amenable to this style
of interpretation, but it is likely that we will have to abandon \emph{some} of our existing ways
of talking about particles to accommodate this mathematics.

It is also reiterated here, following \cite{MorganWLQF}, that the positive spectrum condition
on the energy, which has been so much part of the quantum field theoretical landscape, should
be deprecated, because energy (and as well energy density) is unobservable, infinite, and
nonlocal.
If we think of the random field that is the classical equivalent of a given quantum field,
taking $[a_f,a_g^\dagger]=(g,f)+(f,g)$ so that the commutator is real and $[\hat\phi_f,\hat\phi_g]=0$
for all test functions, it is clear that we are discussing an essentially fractal structure,
for which differentiation and energy density at a point are undefined.
From a proper mathematical perspective, we should consider only finite local observables.
We have accepted renormalization formalisms that manage infinities only in lack of a finite
alternative, a basis for which this paper and its precursor provide.

The method of section \ref{FieldDeformationII} is perhaps more significant mathematically than the
methods of sections \ref{VacuumDisplacement} and \ref{FieldDeformationI}, insofar as the quantum
field observables of section \ref{FieldDeformationII} satisfy modified commutation relations, in
common with the methods for constructing nonlinear quantum fields that are presented in
\cite{MorganWLQF}.
However, quantum theory somewhat exaggerates the importance of commutation relations between quantum
mechanically ideal measurement devices --- the trivial commutation relations of classically ideal
measurement devices can give a description of experiments that is equally empirically
adequate\cite{MorganVaxjo,MorganBellRF}, and ideal measurement devices between the quantum and
the classical can also be used as points of reference\cite{MorganM}.

Physics emphasizes a commitment to observed statistics, which present essentially uncontroversial
lists of numbers, but it is far more difficult to describe what we believe we have measured
than the statistics and the lists of numbers themselves.
It might be said, for example, that ``we have measured the momentum of a particle'', and cite a
list of times and places where devices triggered, ignoring the delicate questions of (1) whether
there is any such thing as ``a particle'', (2) whether a particle can be said to have any
well-defined properties at all, and (3) whether particles have ``momentum'' in particular.
It makes sense to describe a measurement in such a way, because it forms a significant part of a
coordinatization of the measurement that is good enough for the experiment and its results to be
reproduced, but an alternative conceptualization can have a radical effect on our understanding.

\end{document}